# PepINVENT: Generative peptide design beyond the natural amino acids

Gökçe Geylan[ab*], Jon Paul Janet[a], Alessandro Tibo[a], Jiazhen He[a], Atanas Patronov[c], Mikhail Kabeshov[a], Florian David[b], Werngard Czechtizky[d], Ola Engkvist[ae] and Leonardo De Maria[d]

*a. Molecular AI, Discovery Sciences, BioPharmaceuticals R&D, AstraZeneca, Gothenburg, Sweden.*
*b. Division of Systems and Synthetic Biology, Department of Life Sciences, Chalmers University of Technology, Gothenburg, Sweden.*
*c. Quantitative Biology, Discovery Sciences, BioPharmaceuticals R&D AstraZeneca, Gothenburg, Sweden*
*d. Medicinal Chemistry, Research and Early Development, Respiratory & Immunology, BioPharmaceuticals R&D, AstraZeneca, Gothenburg, Sweden*
*e. Department of Computer Science and Engineering, Chalmers University of Technology and University of Gothenburg, Gothenburg, Sweden*
* Email: gokce.geylan@astrazeneca.com

## Abstract

Peptides play a crucial role in the drug design and discovery whether as a therapeutic modality or a delivery agent. Non-natural amino acids (NNAAs) have been used to enhance the peptide properties from binding affinity, plasma stability to permeability. Incorporating novel NNAAs facilitates the design of more effective peptides with improved properties. The generative models used in the field, have focused on navigating the peptide sequence space. The sequence space is formed by combinations of a predefined set of amino acids. However, there is still a need for a tool to explore the peptide landscape beyond this enumerated space to unlock and effectively incorporate *de novo* design of new amino acids. To thoroughly explore the theoretical chemical space of the peptides, we present PepINVENT, a novel generative AI-based tool as an extension to the small molecule molecular design platform, REINVENT. PepINVENT navigates the vast space of natural and non-natural amino acids to propose valid, novel, and diverse peptide designs. The generative model can serve as a central tool for peptide-related tasks, as it was not trained on peptides with specific properties or topologies. The prior was trained to understand the granularity of peptides and to design amino acids for filling the masked positions within a peptide. PepINVENT coupled with reinforcement learning enables the goal-oriented design of peptides using its chemistry-informed generative capabilities. This study demonstrates PepINVENT's ability to explore the peptide space with unique and novel designs, and its capacity for property optimization in the context of therapeutically relevant peptides. Our tool can be employed for multi-parameter learning objectives, peptidomimetics, lead optimization, and variety of other tasks within the peptide domain.

## Introduction

Peptides occupy larger surface area than small molecules and smaller than proteins, occupy a unique space in drug development.[1] With their high specificity, affinity, and low toxicity, peptides have been receiving increased attention from the drug discovery and development area.[2] Their ability to bind to larger surface areas allows them to effectively target protein pockets, shallow clefts on the protein surfaces and protein-protein interaction interfaces, including those deemed to be undruggable by small molecule drugs.[1,3] With the 20 proteinogenic amino acids serving as the building blocks, the peptidic enumerated space expands exponentially. This space covers a vast combinatorial range of $20^L$ variations, where L represents the length of the peptide sequence.[4] However, in nature, it is common for peptides to be modified and diverge from this space with examples such as post-translational modifications in cellular processes or bacteria and fungi synthesizing non-proteinogenic amino acids.[1] These modified peptides impact the survival of many organisms by enabling the signaling and regulation of metabolic pathways. Additionally, they improve potency for protection mechanisms like producing neurotoxins.[5]

In many peptide drug projects, a peptide hit is identified through large library screenings navigating the enumerated space.[6] The integration of non-natural amino acids (NNAAs), amino acids not encoded by DNA, offers a compelling opportunity to improve the physicochemical and pharmacokinetic profile of peptides in the hit-to-lead development. This includes enhancing metabolic stability, binding affinity, or cell permeability.[5,6] The incorporation of NNAAs enables researchers to access even a broader and more diverse chemical space. Considering only the α-amino acid space, each side chain is chosen from a space similar to that of small molecules. Exploring this uncharted space has transformed peptide therapeutics, allowing further refinement of drug designs for better target specificity of both established and novel biological activities.

Conventional methods such as display technologies, peptidomimetics, structure-based computational studies, have been instrumental to the progress of peptide therapeutics.[1] While these methods have played a crucial role in peptide design, they are often limited by the natural amino acids.[1] Even though simple modifications are included in this library, such as stereochemical modifications, the reach of their design space still falls short of the potential scale offered by NNAAs. The virtual space presents a significant challenge to create and is always constrained by the capabilities of design-make-test-analyze cycle. Generative models have been employed to accelerate the drug discovery and development process to efficiently navigate the chemical space. Generative capabilities allow de novo design or molecule optimization with desired properties.[7] In the recent years, there have been many generative modelling studies to design peptides with various optimization tasks such as antimicrobial activity, cell penetration, anticancer and immunogenicity.[4] These studies differ by the characteristics of peptides, the representations and the model architectures they explore however, they have a common goal of designing a peptide sequence with a set number of amino acids, typically 20 natural amino acids.[8–12] Grisoni *et al.* used a long short-term memory (LSTM) model, trained on cationic amphipathic peptides, and fine-tuned on known anticancer peptide sequences.[11] The model was later used to design membranolytic anticancer peptides, composed of natural amino acids. The novel peptide sequences were later validated experimentally for anticancer activity.[11] In other applications, the NNAAs were introduced into the building block library to expand the generative model's enumerated space. One example of this was Schissel *et al.* introducing a peptide generator to design peptides for antisense delivery within an enumerated library.[12] Their approach incorporates three unnatural amino acids to the enumerated space of both the generative and the predictive model. A generator-predictor-optimizer loop operates in this expanded repertoire to design enhanced antisense delivery with peptides with lower Arginine content in a diversity-conscious manner. The learning loop, mimicking a directed evolution scenario with the genetic algorithm, was shown to propose peptide designs with the desired properties.[12] These generative model applications enable better access to the chemical space and can provide a greater diversity of designs compared to conventional methods. However, the research within the enumerated set of building blocks restricts the peptides to a sequence-level design. Despite the demonstrated uptake of generative models in navigating the peptide sequence space, there remains a need for a design tool that efficiently optimizes peptides within the fully enabled chemical space.

To address the need for flexible generation of natural amino acids and NNAAs, we introduce PepINVENT tailored for *de novo* peptide design. PepINVENT stems from the REINVENT framework[13,14]. In small molecule realm, the state-of-the-art REINVENT framework utilizes reinforcement learning (RL) with a generative model trained on the chemical language, Simplified Molecular Input Line Entry System (SMILES)[15], to design *de novo* molecules through a multi-parameter optimization (MPO) scenario.[13] Analogous to REINVENT, PepINVENT is an open-source framework consisting of a chemistry-aware pretrained generative model coupled with RL. The framework facilitates the generation of novel NNAAs and diverse peptide topologies to design novel peptides. Inspired by the translation process of proteins and peptides in ribosomes, PepINVENT learns the peptide space on a per amino acid basis and preserves the intricate granularity of the peptide structure. As the generative model proposes amino acids, reinforcement learning guides the overall peptide design using a goal-oriented approach. We demonstrate the potential of PepINVENT to accelerate the peptide drug discovery and development pipeline by extending the design capabilities to novel NNAAs. The tool is suitable for *de novo* design, peptidomimetics, lead optimization and/or peptide property optimization tasks. In this work, we illustrate the utility and effectiveness of PepINVENT through a series of experiments, showcasing: i) its navigation within the peptidic chemical space, ii) its capability for the flexible generation of diverse peptide topologies, iii) how it can be used to perform MPO for peptide property optimization, with the example of enhancing the permeability and solubility for cyclo REV binding protein.

## Methods

### *Training Data Preparation*

Peptide data is scarce, especially when the NNAAs are concerned. The chemical space that can be covered by an enumerated library composed of the known amino acids is rather limited compared to that of small molecules. To overcome this challenge, we generated semi-synthetic peptide data to span a greater and more diverse chemical space. In addition to natural amino acids, NNAAs from the virtual library proposed by Amarasinghe *et al.*[16] were employed to obtain our building block library. The virtual library was constructed by identifying reagents from eMolecules that could be utilized as precursors for amino acids in common one-step synthetic approaches. This reaction-based enumeration yields a diverse set of 380,000 readily synthesizable NNAAs in which a representative subset of 10,000 non-natural α-amino acids made publicly available.[17] The generative models were previously shown to be able to effectively navigate much larger chemical spaces compared to the space covered by the training data.[18] Therefore, utilizing the virtual library of amino acids can potentially uncover novel amino acids and in turn peptides beyond the semi-synthetic training data.

CHUCKLES [19] is a representation method that encodes amino acids in atomic-level with Simplified Molecular Input Line Entry System (SMILES)[15]. This representation follows its own standardized SMILES pattern in the monomer-level. The pattern starts with the amino group in the backbone followed by the α-Carbon, the sidechain, and the remaining backbone. This standardized format of N-to-C denotes the carboxyl group as carbonyl when used in a peptide sequence. Therefore, a sequential concatenation of the CHUCKLES strings of amino acids yields a valid SMILES pattern for the peptide, enabling syntactically correct peptide representation. Our building block library was translated to CHUCKLES pattern after removing the charges from the amino acids.

The generation of semi-synthetic peptide data encapsulated a decision scheme for peptide length, topology, NNAA content, and common mutations, i.e. stereoisomerism information and backbone N-methylations. The data scheme begins with the selection of a peptide topology among the options of linear or variations of cyclic, including head-to-tail, sidechain-to-tail or disulfide bridging.

Downstream decisions are made for the predefined number of samples for the query topology. Initially, the number of amino acids in the peptide, or the peptide length, was indicated by selecting a length between 6 to 18 from a normal distribution (Figure 1). Subsequently, the fraction of NNAAs was determined through random sampling from a left-skewed normal distribution, covering the range of [0, 0.3] (Supplementary Figure 1). Although, the range was arbitrarily selected, it was chosen to recognize that generating a high fraction of NNAAs would significantly impact synthetic feasibility of the peptides. Peptide sequences were enforced to contain up to and primarily around 30% NNAAs by the skewed distribution. Therefore, the semi-synthetic data ensured the generative model to encounter diverse building blocks while continuing to learn within the traditional chemical space with the natural amino acids. The chosen fraction was utilized to define the number of natural amino acids and NNAAs needed for the selected size. The determined numbers of amino acids were sampled without replacement from their respective sets.

Sidechain-to-tail cyclic peptides were constructed through a different amino acid scheme. The amino acids contributing to this cyclization were determined by selecting an amino acid containing a primary amine in its sidechain for the cyclization start and randomly selecting an amino acid for the cyclization termination. The amino acid for cyclization start was placed to a random position in the given length, fulfilling the condition of forming a cycle with at least 5 amino acids. Similarly, disulfide bridging was achieved by selecting two amino acids containing a sulfhydryl in their sidechains. In both cases, the remaining positions were filled by sampling the natural and non-natural sets according to the selected fractions. The chosen amino acids, except for those involved in cyclization, were randomly shuffled to mix the natural and non-natural building blocks.

Amino acids selected for a peptide sequence, were preprocessed by a series of modifications, starting with the stereochemical mutations. The scheme follows a similar trend as the amino acid selection. Initially, a fraction of amino acids containing stereochemical mutations was determined through sampling from a left-skewed distribution (Supplementary Figure 1). The chosen fraction could range from 0 to 0.25. This fraction determined the number of amino acids in the peptide to be modified. A random sampling of the amino acids according to this fraction determined the specific amino acids to be modified. To achieve the stereoisomeric modification, a string manipulation of the stereochemical information was implemented. The backbone N-methylation was incorporated into a subset of amino acids by replicating the selection process used for stereochemical modification.

Finally, we conducted a preprocess step to achieve the selected topology. If the topology is linear, the amino acids were concatenated to form the SMILES string with the carboxylic acid of the last amino acid completed (Figure 2). In case of cyclization, the atoms of the amino acids involved in the cyclization were modified to denote the beginning and the end of the ring structure. The amino acids selected to contain the topological information of peptides or for N-methylation was examined to not contain a secondary amine group in the backbone, eg. Proline, for standard cyclization. The distributions for the modification decisions are placed in the Supplementary Figure 1. The generated data contained similar distributions of varying size, sequence, NNAA content and modifications to amino acids for each topology.

A total of 1 million unique peptides were generated in this scheme, comprising of 40% linear sequences and an equal distribution of the remaining topology categories, 20% each. The peptides with varying topology, size, sequence, NNAA content and modifications to amino acids were split into 90% training, 5% validation and 5% test sets with stratification to preserve both the peptide length and topology distributions. To evaluate the performance of the generative model, two test sub-sets were prepared from the test set. The first set consisted of 100 masked peptides from each topological class, totaling to 400 masked peptides. The peptides from each of these classes were selected in a stratified manner based on the peptide length. This set was used to assess the generative model performance. The second set was utilized for assessing if the model understands the topological context of the peptides. This set was created from the test data by taking 10 peptides from each topological category, totaling to 40. The selected peptides with cyclized topologies, had one of the amino acids with the topological information unmasked where the second one was included among the masked positions.

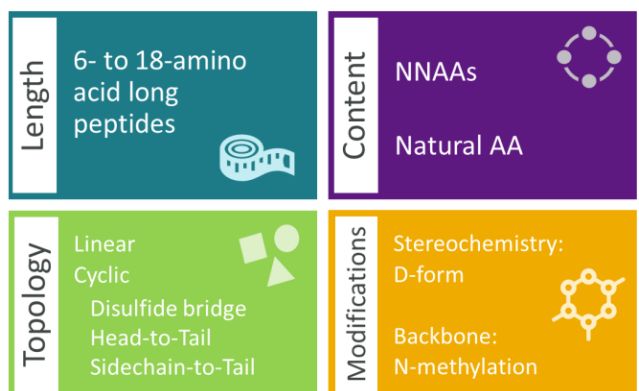

Figure 1. The characteristics of the semi-synthetic data. The peptide data generated to span a range of peptide lengths, different natural-to-non-natural amino acid fractions, peptide topologies and exhibit stereochemical and backbone modifications.

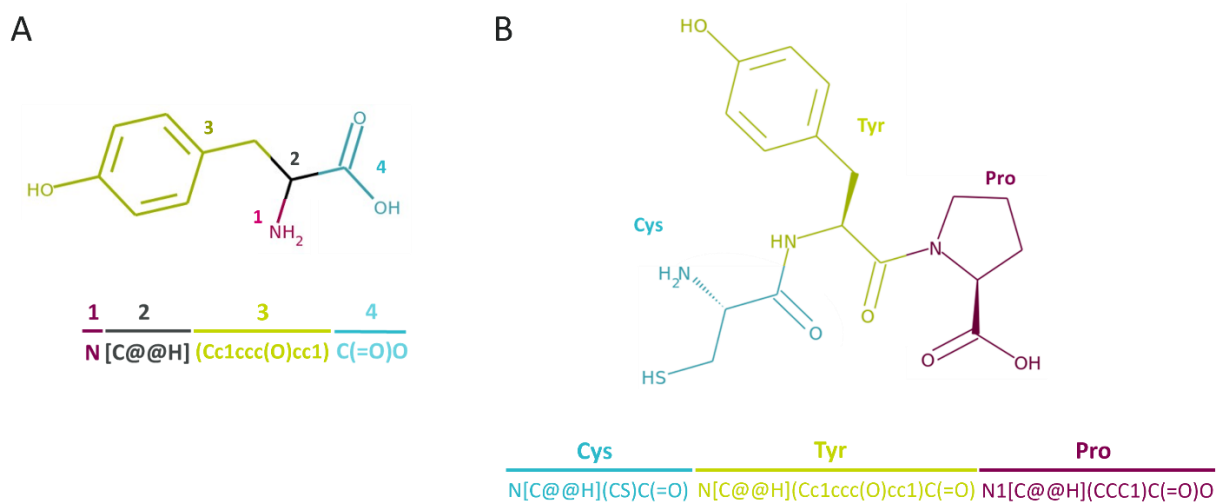

Figure 2. CHUCKLES representation for A) an individual amino acid, Tyrosine (T), and B) a tripeptide, CTP (Cys-Tyr-Pro).

## Pretraining objective

The generative model was designed to propose amino acids for specific positions within a peptide where modifications are desired. We constructed a semi-synthetic dataset consisting of pairs extracted from the peptides. Each peptide was represented by a string, denoted as ρ, which contains the concatenated CHUCKLES of its amino acids, separated by "|". Each pair consists of a source string, $x$, and a target string, $y$. The source string $x$ is formed by replacing certain amino acids in ρ with the special character "?" and moving those amino acids to the target string $y$ (Figure 3). The separator, "|" facilitates a straightforward mapping of the target to source.

The pairs were generated by masking 30% of the amino acids in the peptide. The selection of amino acids to mask involved determining the fractions of natural amino acids and NNAAs. The fraction for natural amino acids was randomly sampled between 0 and 0.5 from a left-skewed distribution, with mean around 0.3 (Supplementary Figure 1). Therefore, the amino acid selection was biased towards more NNAAs overall to prevent overfitting on the natural amino acid patterns. Natural and non-natural amino acids were randomly masked according to their respective assigned fractions. The pretraining objective was defined as proposing a set of amino acids to fill the masked positions of the input peptide. When the number of the generated amino acids equals to the number of masked positions, the generated amino acids, target, were mapped to the source peptide, resulting in the generated peptide.

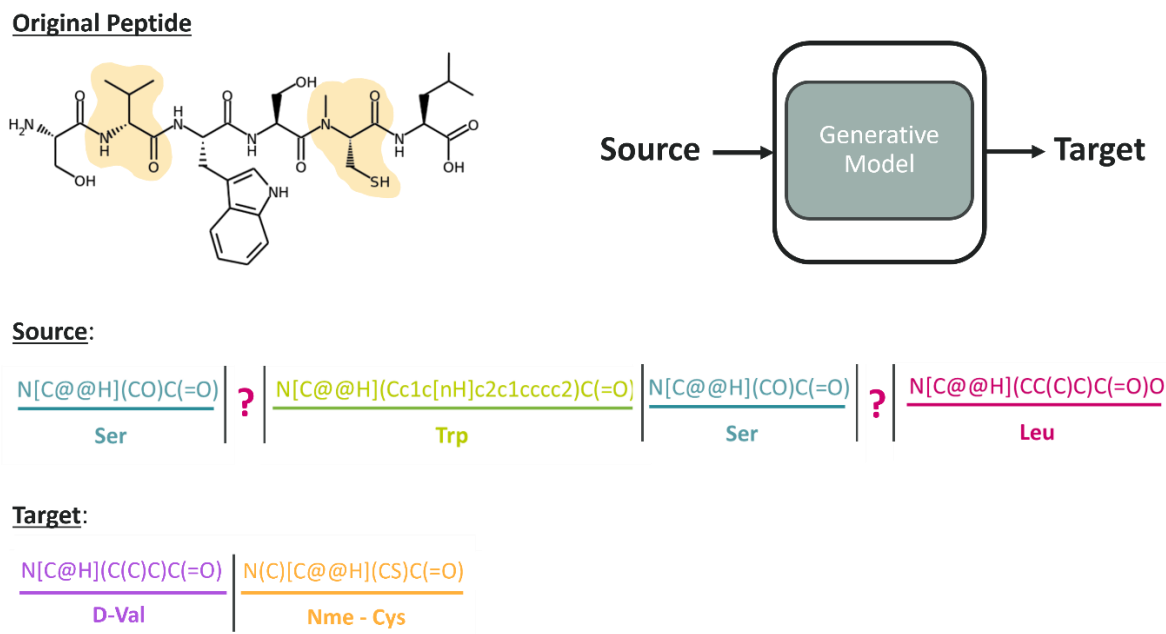

*Figure 3. A source-target pair to illustrate the text-infilling task the conditional generator was trained on. The amino acids at positions 2 and 5 of a 6-mer peptide, denoted as the original peptide, was highlighted as the residues that can be subjected to modification. The source was constructed as the original peptide with positions 2 and 5 masked whereas the target shows the two filler amino acids to complete the masked positions. The model trained on source-target pairs, learns to generate the required number of amino acids to complete the peptide, to follow the CHUCKLES pattern and to understand the complexity of the peptide context.*

## *Model Architecture and Training*

Our generative model uses the same model as the transformer model used in REINVENT[20]. The transformer model consists of an encoder and a decoder. More details about its structure can be found in Supplementary Information 1. The transformer model is denoted by a function $f_\theta$ parametrized by a set of parameters, denoted by $\theta$.

$$f_\theta : \chi \times \chi \to [0,1]^{|V|}$$

The source and target strings were tokenized with a SMILES-based tokenizer and were input to the encoder and decoder during training, respectively. We denote the vocabulary with $V$, i.e. the set of all the possible tokens.[21] $f_\theta$ assigns the probability of the tokens by which the elements of chemical space $\chi$ are represented. [21]From now on we assume the source and target strings: $x, y \in \chi$. The following loss function was used for training the model:

$$NLL(x, y) = -\sum_{t=1}^{T} \log f_\theta(x, y_{0:t-1})[y_t]$$

where $x$ and $y$ are source and target sequence of tokens. $T$ is the length of the output sequence $y$, and $[y_t]$ defines the index of $y_t$, the $t^{th}$ token of $y$.[21] $f_\theta$ computes the probability of the $t^{th}$ token in the target to be generated conditioned on all the previous tokens, $y_{0,...,t-1}$, and x.[21]

The model was trained for 24 epochs on NVIDIA V100 with 32GB. During an epoch all the source-target pairs in the training set are included once with a batch size of 16 and batches shuffled at each epoch. The model was trained following the same strategy and using the same hyperparameters as the original REINVENT transformer model[20], including Adam optimizer with learning rate 0.0001 with 4,000 warmup steps.

Once trained, the model can be used to generate peptides conditioned on proposing amino acids to fill the masked positions of a source peptide by predicting one token at a time. Initially, the decoder processes the start token along with the encoder outputs to sample the next token from the probability distribution over all the tokens in the vocabulary. The generation process iteratively continues by producing the next token from the encoder outputs and all the previous generated tokens until the end token is found or a predefined maximum sequence length, 500, is reached. To allow for the sampling of multiple generated peptides, multinomial sampling or beam search is used.

The model was trained with 900K masked peptides and their filler amino acid pairs. After training, the model can generate the exact number of amino acids required to fill the masked positions in the input peptide. As the peptides were represented with chemical language for strings of amino acids, the model learns the overall peptide language as a composition of individual amino acids. The chemical language also enables generating novel amino acids, simple modifications such as backbone N-methylation and stereochemical mutations.

## *Evaluation Metrics*

The performance of the generative model was assessed on the task encompassing the generation of the correct number of building blocks to fill the masked positions. Any instance with greater or a smaller number of generated amino acids compared to the masked positions were considered as a failure. When a generated peptide contains as many amino acids as the number of masks, the other evaluation metrics are tested. These metrics include:

1. Validity: A generated peptide with a syntactically accurate SMILES that follows the chemical rules such as valency and chemical bonding was categorized as valid, with validity assessed using RDKit[22].
2. Uniqueness: It is defined in multiple levels:
   2.1. Peptide-level uniqueness: The number of unique SMILES strings after the separators are removed and the generated peptide is canonicalized with chirality. As the model generates amino acids to complete an input peptide, the generation of two peptides from the same input might contain the same amino acids in different orders. This makes the two peptides unique but the unique set of amino acids to be duplicated.
   2.2. Amino acid-level uniqueness: This was evaluated in three levels to detect the non-canonical, stereochemical and canonical variability of the generated building blocks, respectively to:

a) String-Level Uniqueness refers to the number of amino acids strings generated being unique by comparing them character by character.
   b) Isomeric SMILES-Level Uniqueness, similarly to the peptide-level uniqueness, is the number of unique amino acids after the SMILES strings are canonicalized while retaining the chirality.
   c) Canonical SMILES-Level Uniqueness is the unique amino acids with canonicalization as the molecules stripped off their stereochemical information. This offers the standardized representation where the uniqueness is ensured by a distinct molecular structure.
3. Novelty: The novelty was calculated by profiling the unique generated amino acids as natural, non-natural and novel. In this case, non-natural refers to the NNAAs utilized to create the semi-synthetic peptide data for model training whereas novel is the NNAAs that are generated by the model that do not exist in the training set.

We also visualized the chemical space of the amino acids to analyze the diversity of novel amino acids from the natural and NNAAs ones. The diversity was illustrated by a t-distributed stochastic neighbor embedding (tSNE). The 1024-bit Morgan fingerprints with `radius=3`, `useChirality=True` and `useCounts=True`[23] computed with `RDKit` v.2024.03.5[22] was projected to 2-dimensional space with `Scikit-learn` v.0.24.2[24]. All the amino acids profiled during the novelty analysis were colored according to their labels.

## Experimental setup

The experiments aim to showcase the generative model's capabilities in navigation of the peptide chemical space and how it can facilitate peptide design with optimized properties.

   1. *Generative model*

The generative model was assessed through a series of experiments using the first test set described in the "Training Data Preparation" section. The sampling was conducted with two sampling methods, a stochastic method, multinomial, and a deterministic method, beam search, with beam size 1000. 1000 sets of filler amino acids were sampled for each masked peptide for both sampling methods. The multinomial sampling was performed in triplicates to ensure reproducibility, and the reported metrics are the average of these runs. The initial evaluation was the task completion, referring to generation the same number of amino acids as the masked positions. The peptides fulfilling this criterion were evaluated with the rest of the evaluation metrics described earlier in the Methods. The results of these metrics were aggregated over the test peptides by calculating the arithmetic mean to determine the overall performance. Additionally, the evaluation metrics were calculated within the peptide topology categories to detect any performance shifts due to topological constraints.

In the next experiment, we tested if the model learned the topological information of the peptides. For example, a macrocyclic peptide includes two amino acids defining the start and end points of the cyclization thereby, establishing the topological arrangement. This experiment aimed to assess whether the model generates amino acids considering the context of the entire peptide. The second test set comprising 40 peptides was used where 1000 filler amino acid sets were sampled for each peptide. If one of the generated amino acids did not complete the topological arrangement, the resulting peptide was considered as an invalid molecule. Therefore, we evaluated the validity per topology for the test peptides.

### 2. *Reinforcement learning*

The search algorithm conceptualized with RL was adapted from the REINVENT's infrastructure [14]. In the RL loop, a user-specific scoring function containing one or more scoring components is used to score the molecules and tune the model to improve the scoring objectives over learning steps. REINVENT framework is informed on the RL setup and the scoring components by a configuration file and produces an RL run accordingly.

The experiments in this section were designed to demonstrate the capabilities of the generative model for peptide property optimization by guiding the generation process through RL. PepINVENT offers peptide-based scoring components and scores the generated peptides after the filler amino acids are mapped to the masked positions of the input peptide. When multiple scoring components are selected, scores from the components are aggregated by either a weighted average or a geometric mean to compute the final score for each peptide in the learning step. As multinomial sampling was employed, the RL experiments were conducted in triplicates to avoid any potential bias. In all the RL experiments, diversity filter with penalty was used to prevent the repetitive generation of the same molecule[25]. In each step of the RL loop, 32 peptides were generated.

The first experiment was to optimize the peptide to a specific topology by constraining the size of the maximum ring. The topological constraint experiments were assessed over 100 steps of RL loop and the average score over the batch were reported across the learning steps. The second experiment was to showcase a practical example where a peptide is designed to be soluble and permeable and have cyclic structure. In this experiment, custom alerts component used to penalize the generation of undesirable patterns. The configuration files that were used to run the RL experiments, could be found in PepINVENT repository.

### 3. *Scoring Components*
**Topological constraints**

As the model learned various topological arrangements, we primarily focused on sampling distinct peptide topologies versus constraining the generation to a specific form. The size of the largest ring in the generated peptide was used to showcase how the generation could be steered towards a specific topology. The scoring component was used in three different optimization scenarios with various score transformations: i) maximize the ring size, ii) sample only head-to-tail or side-chain-to-tail peptides and iii) generate linear peptides. In the Figure 4, the score transformations used in each experimental run were shown. To have a better understanding of the selected score windows, it is important to highlight that a macrocycle is defined as a molecule containing 12 or more atoms in a ring.[26]

Maximizing the ring size was subjected to a sigmoid score transformation within the window of the macrocycle condition, 12 and an arbitrary high number, 60 (Figure 4.A). For sampling head-to-tail or sidechain-to-tail peptides, the upper bound of the score window was reduced to match the typical number of ring atoms in head-to-tail peptides (Figure 4.B). The double sigmoid ensures equal scoring for macrocyclic peptides while heavily penalizing those outside the window. Lastly, the linear peptides were generated by transforming the scores with a reverse sigmoid within a window of 0 to 60, minimizing the ring size (Figure 4.C).

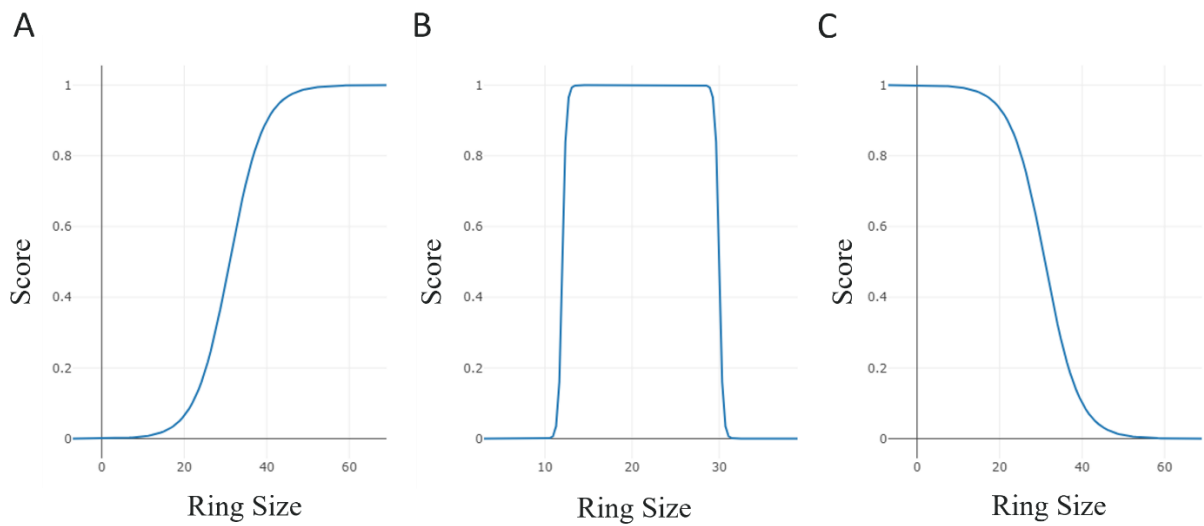

*Figure 4. The score transformations to score the size of the largest ring through A) a sigmoid function fitted between [0,60] to maximize the major cyclization size, B) a double sigmoid function fitted between [0,30] to reward macrocycles smaller or equal to the ring size of head-to-tail cyclization for the given peptide, and C) a reverse sigmoid function fitted between [0,60] to penalize macrocycles and to generate linear peptides.*

To demonstrate the structural flexibility of generation, a 9-mer peptide was generated with positions 1, 2, 4, and 9 masked and the remaining amino acids as Alanine to facilitate the visual distinction of the generated amino acids. Moreover, the input peptide had no prior topological information to enable the generation of any topology. The described input was constructed as: `"?|?|N[C@@H](C)C(=O)|?|N[C@@H](C)C(=O)|N[C@@H](C)C(=O)|N[C@@H](C)C(=O)|N[C@@H](C)C(=O)|?"`

**CAMSOL-PTM intrinsic solubility for peptides**

CAMSOL-PTM is an intrinsic solubility predictor for peptides composed of natural amino acids or NNAAs.[27] The tool is used to assess the changes in solubility upon integration of modified amino acids during post-translational modifications (PTMs). The solubility score of a peptide can be predicted by inputting the NNAAs in SMILES strings and natural ones in their corresponding amino acid letter or SMILES. CAMSOL-PTM was integrated to RL framework as a scoring component where the input molecule is the generated peptide. The calculated score for solubility was transformed with a sigmoid function with limits ranging between 0 and 0.5 and with slope of 0.5.

**Peptide permeability model**

A predictive model for permeability classification of cyclic peptides was built using the practices established in our previous work.[28] XGBoost algorithm was implemented with `xgboost` v.1.7.5 package. trained on the parallel artificial membrane permeability assay (PAMPA) data from the Cyclic Peptide Membrane Permeability Database (CycPeptMPDB)[29]. The training and test sets were split to 90% and 10% respectively with stratification on labels and data sources. The XGBoost model was shown to perform on the test set with a balanced accuracy of 0.78 and Matthew's correlation coefficient of 0.59. The baseline ML model was integrated to RL framework. The scores for peptides were computed as the probability of the Permeable class, using `predict_proba` function. There were no transformations applied to the probability score as the score is in the range of [0, 1].

## Results

To assess the model's generation performance, peptides were sampled from according to input queries from various test sets. The success criteria for the generative model were defined as generating valid, novel, diverse peptides. The assessment of the model performance was followed by a series of experiments aimed to provide more in-depth characterization of the generated peptides at the amino acid-level. Diversity, novelty, and comprehension of the peptide context when generating individual amino acids were considered. These experiments were conducted to demonstrate the model's capacity to navigate the chemical space of both natural amino acids and NNAAs.

Later, peptide optimization through reinforcement learning was explored. RL-based experiments were aimed to demonstrate the flexibility in steering the generation to a specific peptide topology compared to sampling diverse topologies. Lastly, we showcase a practical application in MPO setting to optimize for permeable and soluble cyclic peptides.

### 1. Generative model

To evaluate the performance of our model, we explored the fulfillment of the task objective, the percentage of valid peptides generated by the model, uniqueness, novelty, and diversity of the generated samples on both peptide- and amino acid-level. The model at epoch 10 was chosen as the epoch where both the training and validation loss curves plateau to avoid overfitting to the training data (Supplementary Figure 2).

**Peptide Validity and Uniqueness**

The initial assessment of the model was exploring if the task objective is accomplished. In both sampling methods, there were rare instances of peptides failing the validity due to the model failing to generate as many amino acids as it was required to fill in the masked positions. The maximum failed number of generated amino acids were three out of 1000 corresponding to 0.3%. This case was an outlier in test data considering the mean number of failures being 0.03 (± 0.02). In Table 1, the mean validity of peptides was reported as the percentage of valid molecules over 1000 samples. The validity was further partitioned to distinct topologies that were in the test set to detect whether there is any inflated bias coming from specific molecular structures. More than 99% and 98% of peptides were unique in total, respectively for beam search and multinomial sampling methods. Both methods showed similarly high validity profiles across different topologies but had a slightly lower validity for the sidechain-to-tail bridged peptides (Table 1).

Next, we explored the uniqueness of the generated peptides for both sampling methods. The beam search is deterministic therefore, generates unique strings. However, this does not guarantee that the SMILES representation translates to a unique molecule. Our generative model almost always generates chemically unique peptides, >99% (Table 1). A similar profile was observed with the multinomial sampling with higher fluctuations in non-linear peptide topologies as these were reported with higher standard deviations (Table 1). The peptides with disulfide bridges were harder to diversify with multinomial. This stemmed from the training set containing limited number of amino acids with sulfur in their sidechain. Therefore, the specific topological constraint of having a disulfide substructure in the peptide molecule was harder to learn compared to other topologies. After establishing that our generative model was producing valid and unique molecules across various peptide topologies, the next step was to characterize the building blocks proposed by the model.

Table 1. The validity and uniqueness of the peptides generated by the beam search and multinomial sampling methods. The reported evaluation metrics were further categorized by the peptide topology and reported as the average (± standard deviation) over the 400 test peptides.

| Metric | Sampling Method | Total | Linear | Head-to-Tail | Disulfide Bridge | Sidechain-to-Tail |
|---|---|---|---|---|---|---|
| Peptide Validity (%) | Beam search | 99 (± 7) | 100 (± 0) | 100 (± 0) | 100 (± 2) | 98 (± 14) |
|  | Multinomial | 98 (± 2) | 98 (± 1) | 98 (± 1) | 99 (± 1) | 97 (± 4) |
| Peptide Uniqueness (%) | Beam search | 100 (± 0) | 100 (± 0) | 100 (± 0) | 100 (± 0) | 100 (± 0) |
|  | Multinomial | 98 (± 8) | 100 (± 0) | 99 (± 7) | 94 (± 14) | 99 (± 2) |

**Uniqueness, Novelty and Diversity of Amino Acids**

The unique, novel and diverse amino acids play a central role in assessing the extent of the building block chemical space explored by the generative model and the potential of generating more diverse peptides. In the generation process, multiple amino acids are generated and mapped back to the masked positions of an input peptide in order. Generating sets of the same amino acids in different orders results in distinct peptides, highlighting a potential bottleneck in achieving diversity. To evaluate the model's generative capabilities, we assessed uniqueness at the building block level using generated strings, canonicalized isomeric SMILES, and canonicalized SMILES. In this analysis, we were primarily interested in the number of building blocks generated to propose amino acids for a single input. The uniqueness could vary with many factors such as the number of masks, the amino acid content of input peptide and the topology if the input contains an incomplete ring structure as shown in Table 1. Therefore, the average uniqueness was averaged over the samples generated from the test set. The amino acids had very similar numbers of string level and isomeric SMILES level unique instances, demonstrating that CHUCKLES representation facilitated a standardized format for the amino acids. Both sampling methods showed a decrease in the unique number of amino acids when the chirality information is removed (Figure 5). This underscores the role of stereochemistry in enhancing the diversity of the chemical space and as one of the modification options for the model. The multinomial sampling method notably resulted in significantly larger number of unique amino acids, more than 1400, compared to the beam search, close to 200 (Figure 5). This difference highlights the influence of the sampling strategy when exploring the chemical space and multinomial sampling enabling a broader exploration and diversity of the chemical space. Considering the generative models typically trained on a set of 20 natural amino acids, the PepINVENT model successfully expanded the building block chemical space. On average beam search and multinomial sampling methods respectively resulted in 10- and 70-fold expansion that of the traditional amino acid space for a single peptide input.

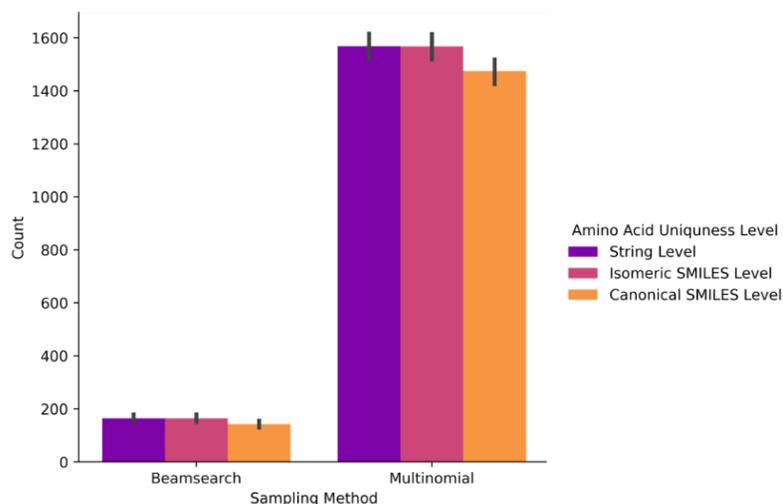

*Figure 5. The three levels of uniqueness for amino acids obtained on average from sampling 1000 peptides for each of the 400 peptides from the test set with beam search, with beam size of 1000 and multinomial sampling methods. The number of unique amino acids include any natural or non-natural amino acids.*

We defined the last step of the uniqueness analysis as the categorization of the type of amino acids generated as detailed in the Methods. In this step, we investigated how the unique amino acids were distributed to the groups of natural, non-natural from training set and novel (Figure 6). When a masked peptide was queried, the entire set of 20 natural amino acid was generally proposed during multinomial sampling. This demonstrates that the generative model considers proposing natural amino acids and not only explore the non-natural space. The NNAAs from the training set were also proposed as the learned building blocks. These NNAAs were more frequently proposed compared to the novel ones. Moreover, in the canonical SMILES-level uniqueness, there was an increase the average of the non-naturals while a decrease in the novel amino acids. This highlights once again the contribution of stereochemical modifications to the diversity. The novel amino acids, at the canonical SMILES-level, were generated with a significant number of options, averaging around 200 and offering as many as 1200 amino acids for a single peptide query.

The drastic difference of the number of amino acids between the two sampling methods arise from how the model learned the amino acid patterns. When a set of amino acids is generated in different orders, it can result in distinct peptides, even though the constituent amino acids are shuffled. In addition, the training set contains more natural amino acids than its non-natural counterpart and some sidechain fragments are frequent among non-natural amino acids. The CHUCKLES pattern for these amino acids and substructures are learned by the model. Considering these two points, the beam search sampling may result in oversampling of natural amino acids and frequent sidechain patterns, the most probable patterns in the training set, compared to multinomial. However, beam search maintains the peptide's uniqueness through positional rearrangements of the amino acids within the peptide. This shows that the model tends to prioritize suggesting natural amino acids initially before venturing into the space of NNAAs. Moreover, shuffling the order of the amino acids illustrates the model's approach in addressing the assigned task by generating a variety of amino acid combinations in a combinatorial fashion. While this may be the case for beam search, the probabilistic nature of multinomial explores the chemical space more freely while preserving the understanding of peptides as a combination of amino acids. Hence, the peptide-level diversity expands into a high-dimensional space that is incomparably broader than the conventional sequence space, requiring a strategic navigation.

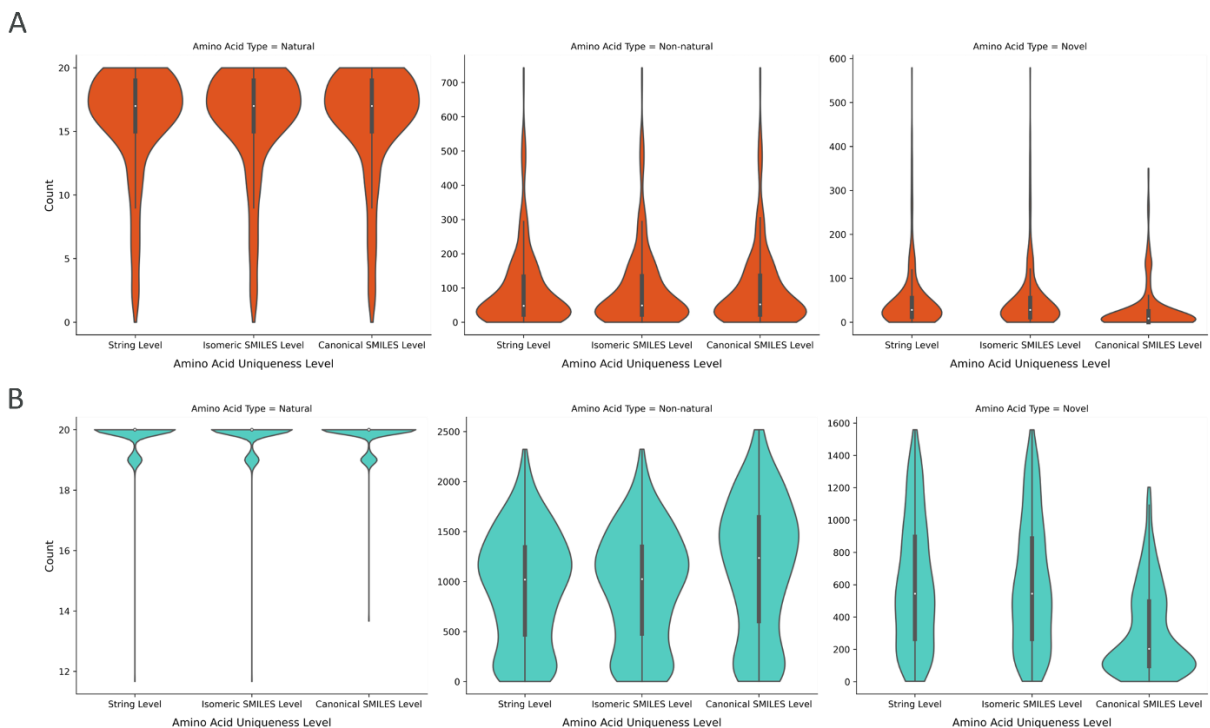

*Figure 6. The distributions of the unique amino acids for three levels of uniqueness and categorized by the type of amino acids: natural, non-natural from the training set and novel. The projection of uniqueness on the types of amino acids were plotted for the output from A) beam search, and B) multinomial sampling methods.*

Lastly, the diversity of the amino acids proposed by the model were analyzed by visualizing the chemical space. As the NNAAs in our training set were already shown to cover a large chemical space[16], the diversity analysis also described the chemical space that was shown to the model with the training set. The extracted amino acids were canonicalized with isomeric information and covered all the natural amino acids and 10,000 NNAAs in the training set. Moreover, 91,826 novel amino acids were generated to propose amino acids for peptides in the test set. The dimensionality reduction plot showed similar coverage for novel and NNAAs, indicating that the novel amino acids were indeed proposed from the learned space (Figure 7). The visualization was plotted with the canonicalized isomeric SMILES since one of the features that the generative model offers is introducing stereochemical modifications.

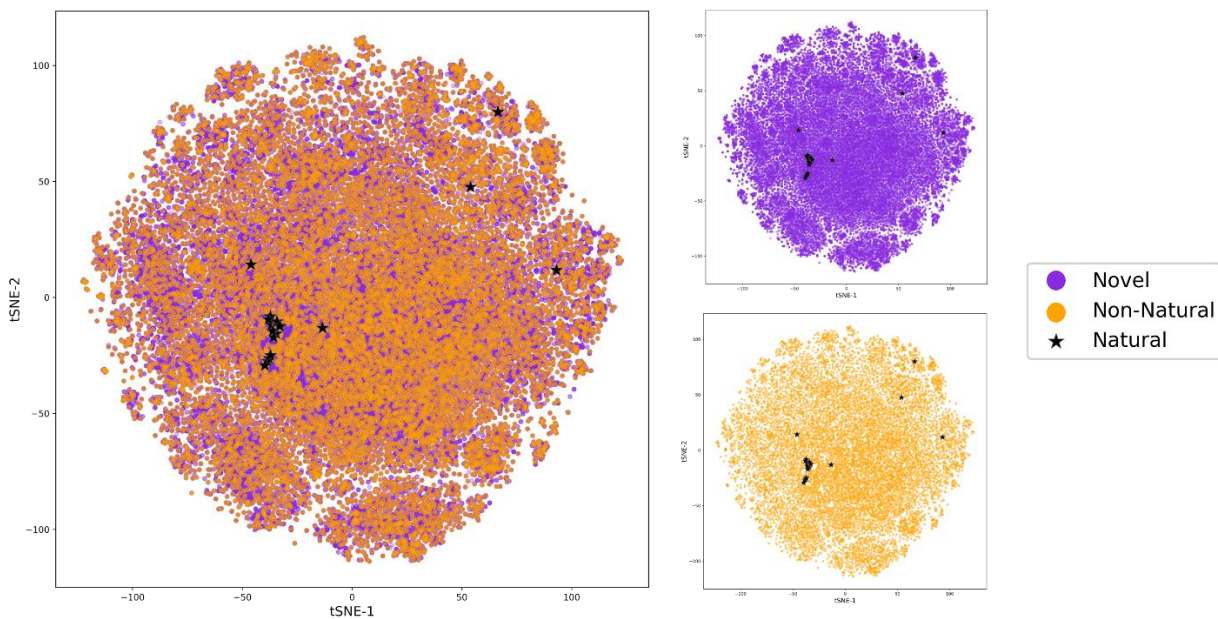

*Figure 7. The chemical space visualization plotted with dimensionality reduction method of t-SNE. The amino acids are colored by the amino acid categories defined as natural, non-natural from the training set and novel amino acids. The chemical space also illustrated separately for novel, around 92K NNAAs, and the 10K NNAAs from the training set, all recovered in sampling, to highlight the overlapping trends between these types of amino acids.*

**Learning the Topological Context**

In the previous sections, we have established that the model can generate a specific number of valid, unique and novel amino acids to complete a masked peptide. We further assessed the model's understanding of the peptide topology by evaluating if there are potential limitations to validity when an uncompleted peptide topology is presented to the generative model. In such cases, there were no significant changes in validity and the model generated amino acids to complete the topology (Table 2). For example, when there is an amino acid with the indication of participating in a disulfide bridge, the model would generate one of the amino acids to fulfill the structural query. In this case, one amino acid would contain a sulfur in its sidechain and a SMILES token symbolizing the bridging would be incorporated to the CHUCKLES representation. This highlights the model's ability to comprehend the input peptide not only as a query for a specific number of amino acids but also with the peptide context.

*Table 2. The percentage of validity of the generated peptides were reported as the means and standard deviations in the triplicate runs for distinct topologies and the overall test set.*

| Topology | Validity (%) |
|---|---|
| All Topologies | 98.3 (± 6.9) |
| Linear | 99.9 (± 0.5) |
| Head-to-Tail | 96.1 (± 12.2) |
| Disulfide Bridge | 97.9 (± 6.5) |
| Sidechain-to-Tail | 99.4 (± 1.8) |

## *2. Reinforcement Learning*

In this study, we conducted reinforcement learning experiments to show the capability of steering the generation process to constrain the generation to a specific peptide, and to optimize peptide properties in a MPO scenario.

**Scenario 1: Generating a peptide topology of interest**

We demonstrate a single scoring component, the maximum ring size of the peptide, effectively driving the generation to a specific peptide topology. We conducted simulation of multiple RL-based runs using different score transformations preferring distinct topologies. The learning processes were visualized focusing on the change of the maximum ring size of the batch of peptides generated in every step (Figure 8). Additionally, the learning processes showing the triplicate RL runs can be found in Supplementary Figure 3. The 9-mer head-to-tail cyclized peptide queried in this scenario, is likely to have 27 atoms in its backbone if all the amino acids are stitched together by a peptide bond. Sigmoid score transformation favoring the larger sizes, showed an increase of the average ring size to a range between 30 and 35 (Figure 8.A). This implies that the generated peptides have the tendency to contain larger rings than a head-to-tail cyclized peptide. A peptide containing a disulfide bridge connecting its first and the last amino acid can be an instance for this. Such peptide would include all the backbone atoms and the side-chains atoms in between the backbone and the Sulfur attachment points in its largest ring. In line with this information, the generated topologies were observed to be more representative of peptides with disulfide bridges. Moreover, PepINVENT generated some bicyclic peptides which is a novel topology compared to the training set, highlighting the navigation of the model in the unseen chemical space (Supplementary Figure 4).

In the next RL case, we have biased the scoring to have the best scores throughout the range of the macrocycle condition to the head-to-tail cyclization condition. The learning was limited to these topologies as the average ring size increased until it reached the pre-set threshold of the score transformation (Figure 8.B). Moreover, the broader score range of the steps compared to the disulfide case demonstrates the fluctuations of generating both topologies (Supplementary Figure 4). Lastly, we flip our initial objective to favor lower ring sizes or in other words, linear peptides generation in the RL run. Once again, RL was able to steer the generation to linear peptides as the macrocycles are penalized with lower scores (Figure 8.C). The generation of linear peptides also did not affect sampling heterocycles in the sidechains (Supplementary Figure 4). The preservation of such substructures ensures the diversity of sidechains while conforming to the desired topology. In all the RL cases, the validities of the batch of peptides in the exploration and exploitation stages were generally above 90% and 95%, respectively (Figure 8.D-F). Typically, at approximately 40 learning steps, the objective is reached, marking the transition from exploration of the peptidic chemical space to exploitation of the targeted space with the desired characteristics. Achieving the objective under 50 steps, PepINVENT demonstrated high flexibility in transitioning between diverse topologies.

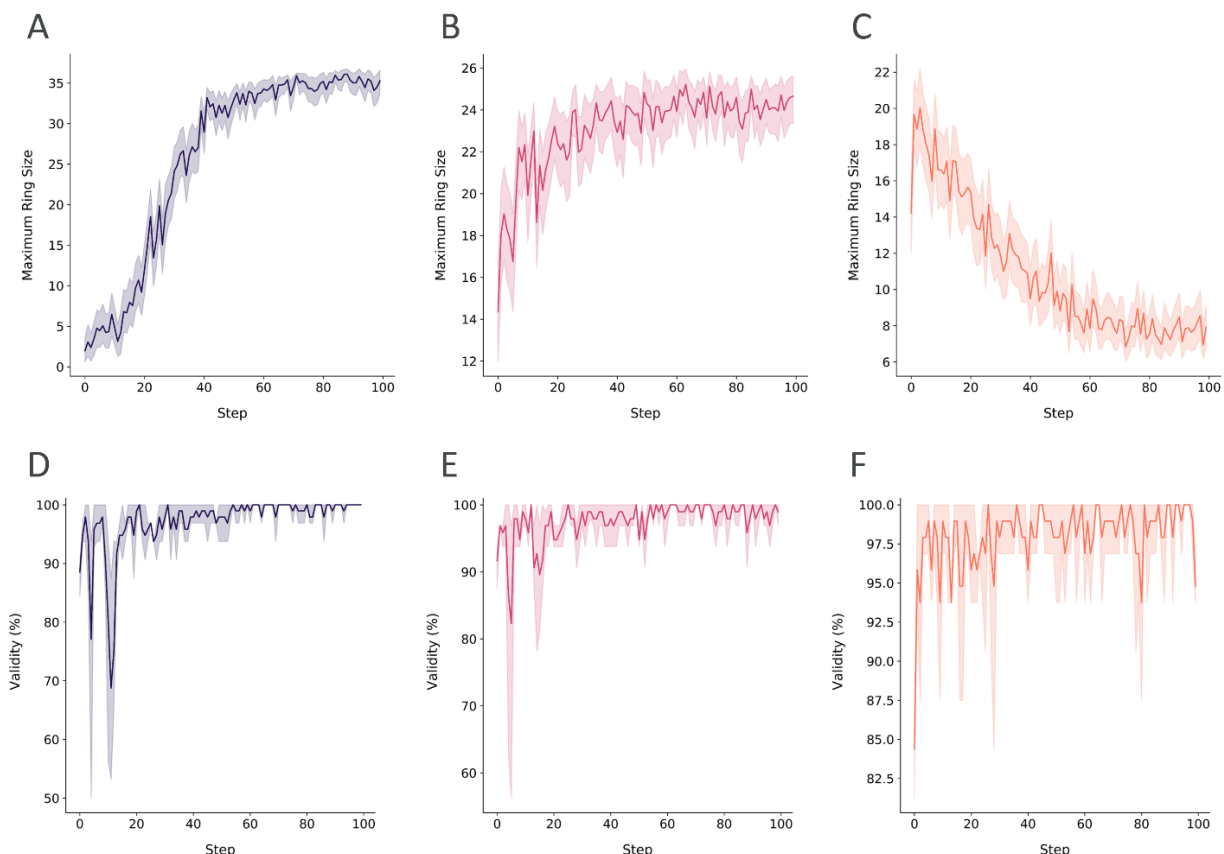

*Figure 8. The scoring component, the ring size of the largest ring, (A-C) and the percentage of valid peptides generated (D-F) over the learning steps in RL runs was plotted. The RL runs were set to optimize the peptide generation for maximizing the ring size (purple), preferring macrocycles with an upper limit of ring size (red), and minimizing the ring size (orange). In every learning step, the average ring size of the batch was plotted with the 95% confidence interval.*

**Scenario 2: Generating soluble and permeable cyclic peptides**

In this application example, we target a therapeutically relevant peptide known as the Rev-binding peptide (RBP). Rev protein, a human immunodeficiency virus (HIV) protein, plays an essential role in regulating the transport of mRNA from nucleus to cytoplasm in the infected cells. The mRNA transport is a key component of the viral replication as it allows for the translation of viral products and facilitates viral assembly. Given its influence over the HIV lifecycle, Rev protein has been a target for antiviral therapy.[30] RBP, with the sequence of YPAASYR, was identified as a potential inhibitor of the Rev protein. The peptide was discovered from the structural analysis of

Rev-antibody complex, corresponding to the paratope of the antibody.[31–33] Wu *et al.* transformed this sequence into a cyclic peptide by extending the sequence with two Glycines and head-to-tail cyclization. The RBP was impermeable to cell membrane.[32] Alanines in the peptide were modified to cyclo-Alanines to enhance the permeability of RBP while retaining its bioactivity.[32]

Inspired by this study, we demonstrate how RBP can be modified to improve permeability and solubility. The amino acids that were shown to not have a major impact on the bioactivity in Wu *et al.*'s study, were masked in RBP. These included the Alanines that were previously shown to be modified without a significant change in the bioactivity and Glycines that were incorporated solely for the cyclization. Over the learning steps, the RBP was modified by PepINVENT in a multi-parameter objective (MPO) scenario to propose new RBP designs by generating sets of amino acids. In the MPO, the RBP was optimized for cell permeability through scoring the designs with a peptide permeability model while maintaining cyclic structure by Topological Constraint scoring component. The undesirable substructures were penalized by Custom Alert scoring component and we tracked all the learning progress of the individual components as well as the final score. The learning process over a 1000-step RL run was tracked with the four scoring components mainly (Figure 9). The maximizing the ring size over the peptide stabilized cyclization as the unmasked head amino acid contained the topological information (Figure 9). The custom alerts allowed the generation to maintain consistency with the chemical relevance of the sidechains (Figure 9). As these components defined the targeted peptidic space, components scoring the permeability and the solubility specified the property optimization objectives.

The intricate interplay between solubility and permeability have been explored by many studies.[33] While there are no explicit design guidelines for peptides, it is understood that cell permeable peptides must demonstrate a certain level of solubility in order to unlock the degree of conformational flexibility necessary for passive permeability.[34] The dynamic conformational shifts enable peptides to adapt to both aqueous and cell membrane's lipophilic media.[35] Therefore designing permeable and soluble peptides requires complex design strategies. PepINVENT was demonstrated to navigate the chemical space to propose design ideas balancing solubility and permeability. According to the learning progress in the RL runs with RBP, the solubility component was learned first, in the first 100 steps, followed by improving the permeability in the soluble peptide space (Figure 9.B-C). This was also observed by a steep increase proceeded by a steady improvement in the aggregated score computed by the geometric mean of four scoring components (Figure 9.G). Over the learning steps, the proposed peptides showed high solubility and permeability, while permeability was harder to optimize (Figure 10). Heterocycle incorporation to the backbone was one of the preferred design ideas for MPO, which is a common peptide modification for permeability (Figure 10).[33] Lipophilicity was also reported as the Wildman-Crippen LogP value during the runs to visualize the shifts between solubility and this fragment-based solubility scorer (Figure 9.E).[36] PepINVENT effectively sampled new cyclic peptides that are soluble and permeable while preserving the previously demonstrated high validity of the generative model.

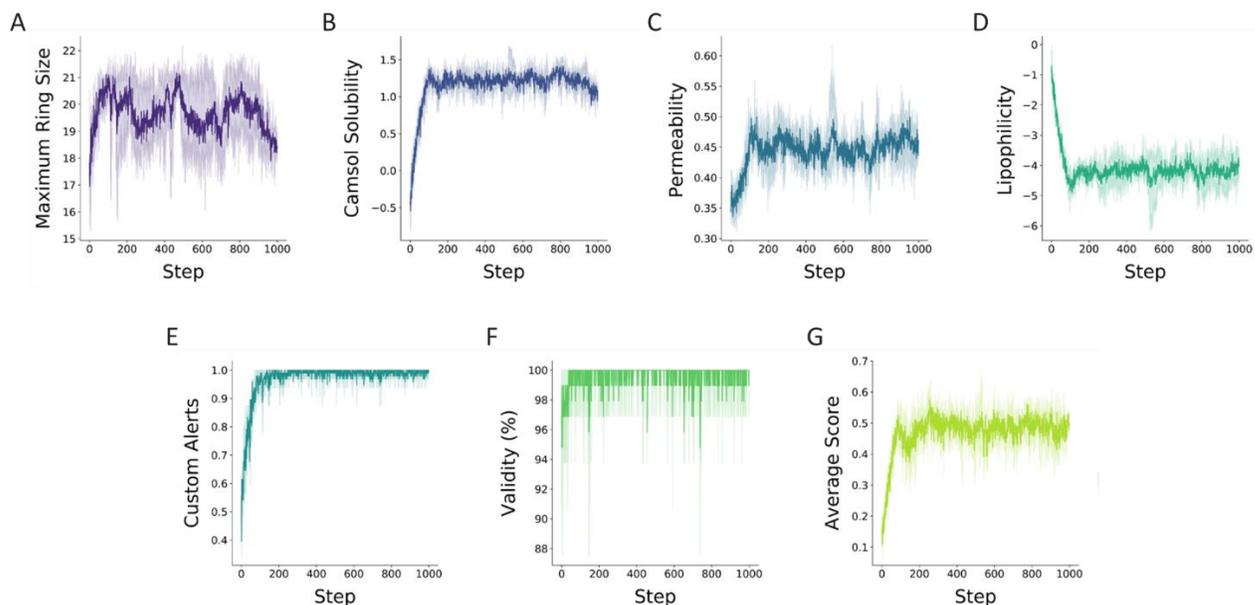

*Figure 9. Progress of the scoring components of the MPO task during RL runs to design soluble and permeable cyclic Rev-binding peptide designs. The MPO entailed A) topological constraint aiming to generate cyclic peptides, B) solubility measured with CAMSOL-PTM intrinsic solubility predictor[27], C) permeability assessed through a classifier for passive permeability of peptides and D) custom alerts penalizing undesirable substructures generally associated with toxicity. During the runs, E) the lipophilicity, F) the validity, and G) the average of the aggregated scores for the batches at each step were tracked for the generated peptides. The plots illustrate the average of triplicate runs and the individual runs were reported in Supplementary Figure 5.*

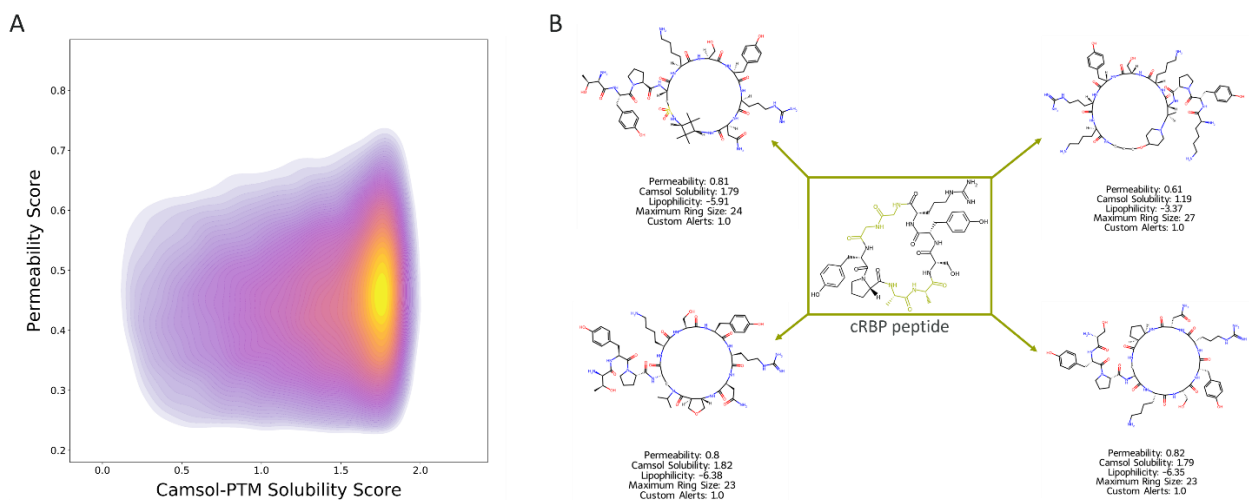

*Figure 10. A) The scatter plot illustrates the distribution of the probability of generated peptides belonging to the permeable class, and the solubility score, colored by the learning steps. There were 6318, 922, and 10 soluble macrocyclic peptides without undesirable substructure with permeability scores above 0.6, 0.7 and 0.8, respectively. B) Examples of generated peptides were reported with MPO components' scores. The generated peptides were positioned around green framed input peptide, cRBP, with the masked amino acid depicted with green color.*

## Discussion

In this work, we introduced PepINVENT as an extension to the de novo design tool for small molecules, REINVENT[13]. PepINVENT is a transformer-based generative model trained to propose peptide designs given a peptide-of-interest. The model performs text infilling by producing novel peptides through generating natural or non-natural amino acids to substitute user-defined positions within the sequence. In contrast to the previous generative methods for peptide design, PepINVENT suggests amino acids with a chemistry-aware generation process, aiming to exploit the full theoretical chemical space of amino acids. Unlocking the atomic-level design for amino acids facilitates the design of novel sidechains, the incorporation of backbone modifications and stereochemical mutations in the peptide. As the uncharted peptide landscape becomes accessible as the design space, the potential of peptide therapeutics to address previously unmet needs expand significantly.

The scarcity of the publicly available peptides and the known NNAAs being significantly limited compared to the theoretical amino acid space pose a challenge for obtaining large data. Because of these data limitations, generative models in the field are typically customized for topology- or property-specific tasks.[8–12] Our generative model was trained on semi-synthetic peptide data composed of natural amino acids and readily synthesizable NNAAs from a large virtual library[16]. The semi-synthetic data was not generated based on any property-related patterns or distinct topologies. The conditional aspect of the generator is defined solely by its ability to generate equal number of amino acids to the masked positions. Therefore, PepINVENT could serve as a central tool for optimizing various peptide-related objectives and topologies.

Peptides were represented with the CHUCKLES pattern to our model. CHUCKLES enabled the peptides to be encoded preserving their chemical context while maintaining the sequence order of amino acids. This approach facilitated the translation from sequence to molecule, allowing for both assembly to peptides and extraction of amino acids from them in a standardized format, capturing the N-to-C directionality. The generative model trained with this representation was shown to produce valid and unique peptides with diverse amino acids while preserving the granular structure of peptides. The model comprehends the complex nature of a peptide as a sequence of building blocks that make up a beyond-the-rule-of-5 modality. The generative model goes beyond the building block library in the training set to propose novel NNAAs featuring unique sidechains, various stereochemical and backbone modifications.

We demonstrated the capabilities of PepINVENT framework encompassing the generative model and the RL through two experiments. The first experiment utilized a straightforward physicochemical descriptor, the ring size of the largest ring in the peptide. Various score transformation functions were employed to define different objectives to optimize the largest ring's size with RL. The experiments showed that when a peptide without a specific structural information can be constrained to a desired topology in less than 50 learning steps. The flexibility in guiding the generation towards distinct topologies highlights its flexibility in proposing diverse peptide designs.

In a peptide-based drug discovery project focused on enhancing the peptide affinity, a motif in the original sequence that leads to a certain characteristic, such as permeability, may be preserved. The lead optimization in such case could be carried out on the amino acids that do not participate in this motif. In contrast, amino acids enabling the peptide's binding to a target protein could be characterized as the pharmacophore.[33] The residue-based pharmacophore is preserved while the remaining amino acids are explored for new peptide designs to improve the peptidic properties, such as solubility or oral bioavailability. PepINVENT can selectively optimize the peptide, with respect to the project goals. The latter case was employed to showcase a practical application of PepINVENT in designing a hit peptide when the pharmacophore is defined. The amino acids that are not part of the pharmacophore or modified in other studies without compromising the activity were selected for modification. The proposed peptide designs were optimized to exhibit permeable and soluble features constrained to cyclic designs. The RL-steered generation led to the proposal of novel peptides with the desired properties. This experiment showed how PepINVENT navigates the peptide landscape identifying limited regions within the chemical space that balances the trade-off among multiple properties. PepINVENT could be utilized for multiparameter optimization consisting of physicochemical properties, property predictors and methodologies. Throughout the experiments, the agent consistently avoided invalid amino acid patterns while navigating the peptide landscape, resulting in over 90% validity of the proposed peptides. Proposing novel peptide designs could accelerate peptide-based drug discovery and development projects. The tool can be used for peptide property enhancements, peptidomimetics, lead optimization and many other peptide related tasks for peptides as drug molecules or as conjugates for delivery. In future studies, we aim to demonstrate the impact of the PepINVENT framework in a multi-parameter optimization setting for a real-life application.

## Conclusion

PepINVENT is a robust generative model-based framework for peptide design. The framework provides a platform for optimizing peptides by proposing single or multiple α-amino acids according to the user-preferred design objectives. The generative model can create novel NNAAs by designing new sidechains, modifying the backbone or the stereochemistry amino acids present in the training set. This novelty could accelerate peptide design by traversing the unexplored chemical space of amino acids. The standardized format, CHUCKLES, offers a chemical language for peptides enabling the model to learn the building block chemistry and the peptide context as a structure composed of a chain of amino acids. The diversity, novelty and uniqueness of the peptide designs were established by analyses conducted on model's generation output.

In this study, we also demonstrated PepINVENT's effectiveness in navigating the peptide landscape by showing its adaptability to specific topological constraints and in showcasing its capacity to accommodate multiple topologies. Additionally, PepINVENT was employed in multi-parameter optimization scenario to design soluble and permeable cyclic peptides. Reinforcement learning steered the generation to the design space that contained desired peptides for all the specified components. Thus, our framework demonstrates its capability to propose tailored peptides and could facilitate peptide optimization in real-life applications. PepINVENT is presented as an open-source framework, and as an extension of the de novo small molecule design tool, REINVENT. In this work, we present PepINVENT as a tool that could facilitate peptide-based drug discovery and development by addressing the challenge of proposing peptide designs with novel NNAAs while improving targeted peptidic properties defined through MPO. Future studies will focus on demonstrating the practical utilization of the framework in peptide-based drug optimization settings.

**Data Availability**

The semi-synthetic data used in model training and testing, the generative model, the configuration files for the experiments shown in the results, and the reinforcement learning code will be publicly available upon publication. The CAMSOL-PTM solubility scorer algorithm used in this study is provided as a web-server freely available for the academic users upon registration at https://www-cohsoftware.ch.cam.ac.uk/index.php/camsolptm.

**Author Contributions**

G.G. contributed to the main part of the research and performed the experiments. G.G., L.D.M., A.P. and O.E. designed and conceptualized the project. G.G., J.P.J., A.T., J.H, L.D.M, and O.E. analysed the results and provided feedback. G.G. wrote the manuscript. The authors discussed the results and reviewed the manuscript.

**Conflicts of interest**

There are no conflicts to declare.

**Acknowledgements**

This work has been supported by the Swedish Foundation for Strategic Research (SSF) (Grant ID20-0109) through funding an industrial PhD studentship for GG.

# References

1.  L. Wang, N. Wang, W. Zhang, X. Cheng, Z. Yan, G. Shao, X. Wang, R. Wang and C. Fu, *Signal Transduction and Targeted Therapy 2022 7:1*, 2022, **7**, 1–27.
2.  G. Rossino, E. Marchese, G. Galli, F. Verde, M. Finizio, M. Serra, P. Linciano and S. Collina, *Molecules*, *Molecules*, 2023, DOI:10.3390/MOLECULES28207165.
3.  N. Tsomaia, *Eur J Med Chem*, 2015, **94**, 459–470.
4.  F. Wan, D. Kontogiorgos-Heintz and C. de la Fuente-Nunez, *Digital Discovery*, 2022, DOI:10.1039/D1DD00024A.
5.  Y. Ding, J. P. Ting, J. Liu, S. Al-Azzam, P. Pandya and S. Afshar, *Amino Acids*, 2020, **52**, 1207.
6.  M. Muttenthaler, G. F. King, D. J. Adams and P. F. Alewood, *Nature Reviews Drug Discovery 2021 20:4*, 2021, **20**, 309–325.
7.  X. Zeng, F. Wang, Y. Luo, S. gu Kang, J. Tang, F. C. Lightstone, E. F. Fang, W. Cornell, R. Nussinov and F. Cheng, *Cell Rep Med*, 2022, **3**, 100794.
8.  X. Xu, C. Xu, W. He, L. Wei, H. Li, J. Zhou, R. Zhang, Y. Wang, Y. Xiong and X. Gao, *Bioinformatics*, 2024, DOI:10.1093/BIOINFORMATICS/BTAE364.
9.  Z. Wu, Y. Wu, C. Zhu, X. Wu, S. Zhai, X. Wang, Z. Su and H. Duan, *J Chem Inf Model*, 2023, **63**, 7655–7668.
10. P. Das, T. Sercu, K. Wadhawan, I. Padhi, S. Gehrmann, F. Cipcigan, V. Chenthamarakshan, H. Strobelt, C. dos Santos, P. Y. Chen, Y. Y. Yang, J. P. K. Tan, J. Hedrick, J. Crain and A. Mojsilovic, *Nature Biomedical Engineering 2021 5:6*, 2021, **5**, 613–623.
11. F. Grisoni, C. S. Neuhaus, G. Gabernet, A. T. Müller, J. A. Hiss and G. Schneider, *ChemMedChem*, 2018, **13**, 1300–1302.
12. C. K. Schissel, S. Mohapatra, J. M. Wolfe, C. M. Fadzen, K. Bellovoda, C. L. Wu, J. A. Wood, A. B. Malmberg, A. Loas, R. Gómez-Bombarelli and B. L. Pentelute, *Nature Chemistry 2021 13:10*, 2021, **13**, 992–1000.
13. H. H. Loeffler, J. He, A. Tibo, J. P. Janet, A. Voronov, L. H. Mervin and O. Engkvist, *J Cheminform*, 2024, **16**, 1–16.
14. M. Olivecrona, T. Blaschke, O. Engkvist and H. Chen, *J Cheminform*, 2017, **9**, 1–14.
15. D. Weininger, *J Chem Inf Comput Sci*, 1988, **28**, 31–36.
16. K. N. Amarasinghe, L. De Maria, C. Tyrchan, L. A. Eriksson, J. Sadowski and D. Petrović, *J Chem Inf Model*, 2022, **62**, 2999–3007.
17. Buy Research Compounds | eMolecules, https://www.emolecules.com/.
18. J. Arús-Pous, T. Blaschke, S. Ulander, J. L. Reymond, H. Chen and O. Engkvist, *J Cheminform*, 2019, **11**, 1–14.
19. M. A. Siani, D. Weininger and J. M. Blaney, *J Chem Inf Comput Sci*, 1994, **34**, 588–593.
20. J. He, E. Nittinger, C. Tyrchan, W. Czechtizky, A. Patronov, E. J. Bjerrum and O. Engkvist, *J Cheminform*, 2022, **14**, 1–14.
21. A. Tibo, J. He, J. P. Janet, E. Nittinger and O. Engkvist, *Nature Communications 2024 15:1*, 2024, **15**, 1–12.
22. D. Rogers and M. Hahn, *J Chem Inf Model*, 2010, **50**, 742–754.

23  RDKit, http://www.rdkit.org/.
24  F. Pedregosa, G. Varoquaux, A. Gramfort, V. Michel, B. Thirion, O. Grisel, M. Blondel, P. Prettenhofer, R. Weiss, V. Dubourg, J. Vanderplas, A. Passos, D. Cournapeau, M. Brucher, M. Perrot and É. Duchesnay, *The Journal of Machine Learning Research*, 2011, **12**, 2825–2830.
25  T. Blaschke, J. Arús-Pous, H. Chen, C. Margreitter, C. Tyrchan, O. Engkvist, K. Papadopoulos and A. Patronov, *J Chem Inf Model*, 2020, **60**, 5918–5922.
26  D. Garcia Jimenez, V. Poongavanam and J. Kihlberg, *J Med Chem*, 2023, **66**, 5377–5396.
27  M. Oeller, R. J. D. Kang, H. L. Bolt, A. L. Gomes dos Santos, A. L. Weinmann, A. Nikitidis, P. Zlatoidsky, W. Su, W. Czechtizky, L. De Maria, P. Sormanni and M. Vendruscolo, *Nature Communications 2023 14:1*, 2023, **14**, 1–12.
28  G. Geylan, L. De Maria, O. Engkvist, F. David and U. Norinder, *Digital Discovery*, 2024, DOI:10.1039/D4DD00056K.
29  J. Li, K. Yanagisawa, M. Sugita, T. Fujie, M. Ohue and Y. Akiyama, *J Chem Inf Model*, 2023, DOI:10.1021/ACS.JCIM.2C01573.
30  rev - Protein Rev - Human immunodeficiency virus type 1 group M subtype B (isolate HXB2) (HIV-1) | UniProtKB | UniProt, https://www.uniprot.org/uniprotkb/P04618/entry.
31  S. J. Stahl, N. R. Watts, C. Rader, M. A. DiMattia, R. G. Mage, I. Palmer, J. D. Kaufman, J. M. Grimes, D. I. Stuart, A. C. Steven and P. T. Wingfield, *J Mol Biol*, 2010, **397**, 697–708.
32  H. Wu, G. Mousseau, S. Mediouni, S. T. Valente and T. Kodadek, *Angewandte Chemie International Edition*, 2016, **55**, 12637–12642.
33  L. K. Buckton, M. N. Rahimi and S. R. McAlpine, *Chemistry – A European Journal*, 2021, **27**, 1487–1513.
34  M. Sugita, S. Sugiyama, T. Fujie, Y. Yoshikawa, K. Yanagisawa, M. Ohue and Y. Akiyama, *J Chem Inf Model*, 2021, **61**, 3681–3695.
35  T. A. Ramelot, J. Palmer, G. T. Montelione and G. Bhardwaj, *Curr Opin Struct Biol*, 2023, **80**, 102603.
36  S. A. Wildman and G. M. Crippen, *J Chem Inf Comput Sci*, 1999, **39**, 868–873.

## Supplementary Material

**Supplementary Information 1:**
The additional information on the transformer model architecture trained for the peptide-based generative model is as follows:

- Encoder layers: six identical blocks consisting of a multi head self-attention and position-wise fully connected feed forward network were used.
- Decoder layers: similarly to the encoder layers, the decoder contains 6 layers. Each decoder layer has an additional sub-layer for multi-head attention on the encoder's output.
- Model dimension: the dimension of the input and output vectors was set to 256. This dimension is the same across all layers and serves as the size of the embeddings and the internal representations within the model.
- Number of attention heads: each multi-head attention mechanism is composed of 8 attention heads. Each attention layer splits the model dimension (256) into 8 subspaces of size 16.
- Dropout rate was set to 0.1.

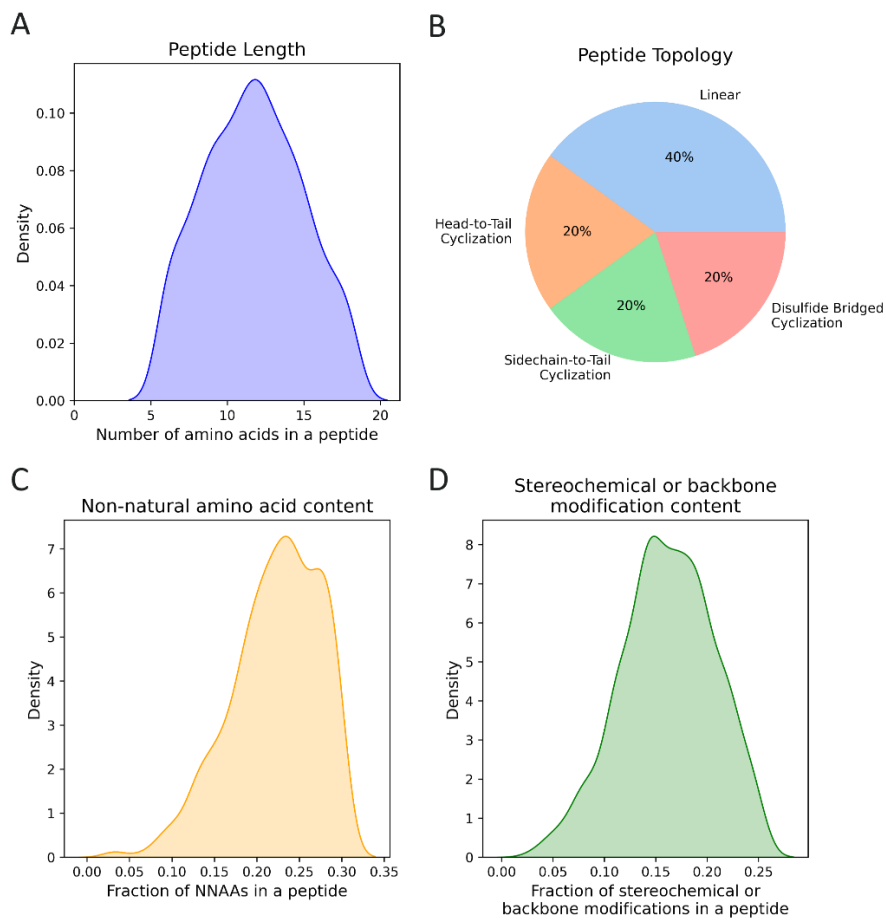

**Supplementary Figure 1.** The distributions of A) the peptide length, B) the topology distribution, C) the fraction of non-natural amino acids (NNAAs) and, D) the fraction of amino acids having N-methylated backbone or stereochemical modification.

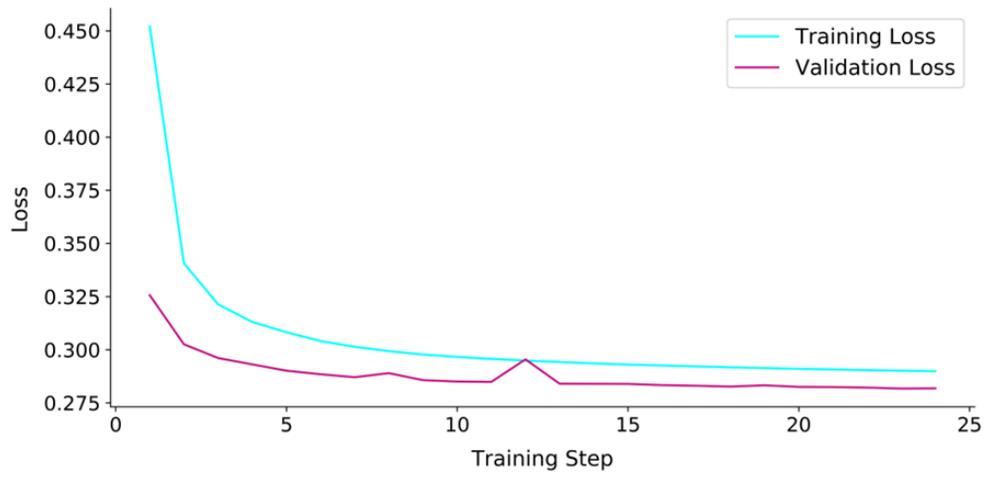

**Supplementary Figure 2.** The training and validation loss curves during the model training. The model state at the epoch 10 in which the losses plateau was selected as the final model.

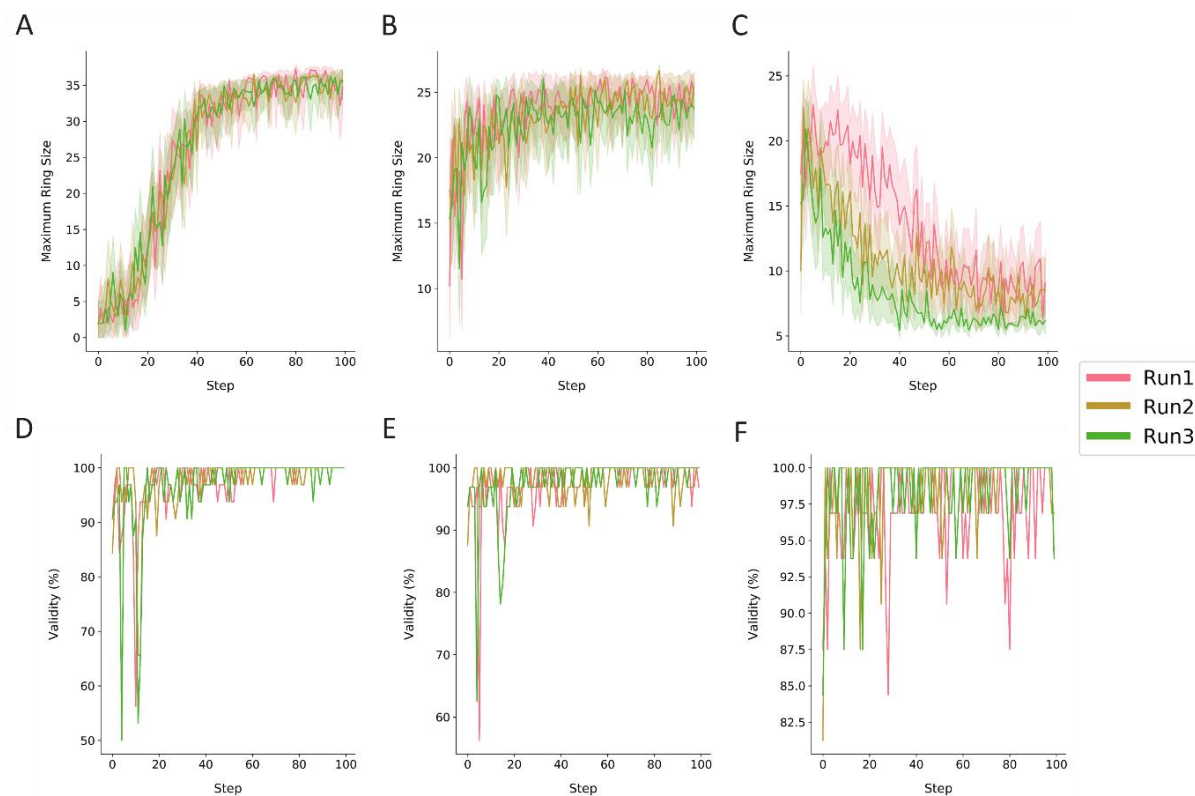

**Supplementary Figure 3.** The scoring component, the ring size of the largest ring, (A-C) and the percentage of valid peptides generated (D-F) over the learning steps in reinforcement learning (RL) runs for Scenario 1: Generating a peptide topology of interest. The evolution of the largest ring size and the validity of the generated peptides were plotted individually for the triplicate runs. The learning objective of A) maximizing the ring size and D) its validity, B) generating any macrocycle with an upper threshold of ring size to yield peptides with sidechain-to-tail or head-to-tail topologies E) its validity, C) minimizing the ring size and F) its validity. In every learning step, the average ring size of the batch was plotted with the 95% confidence interval.

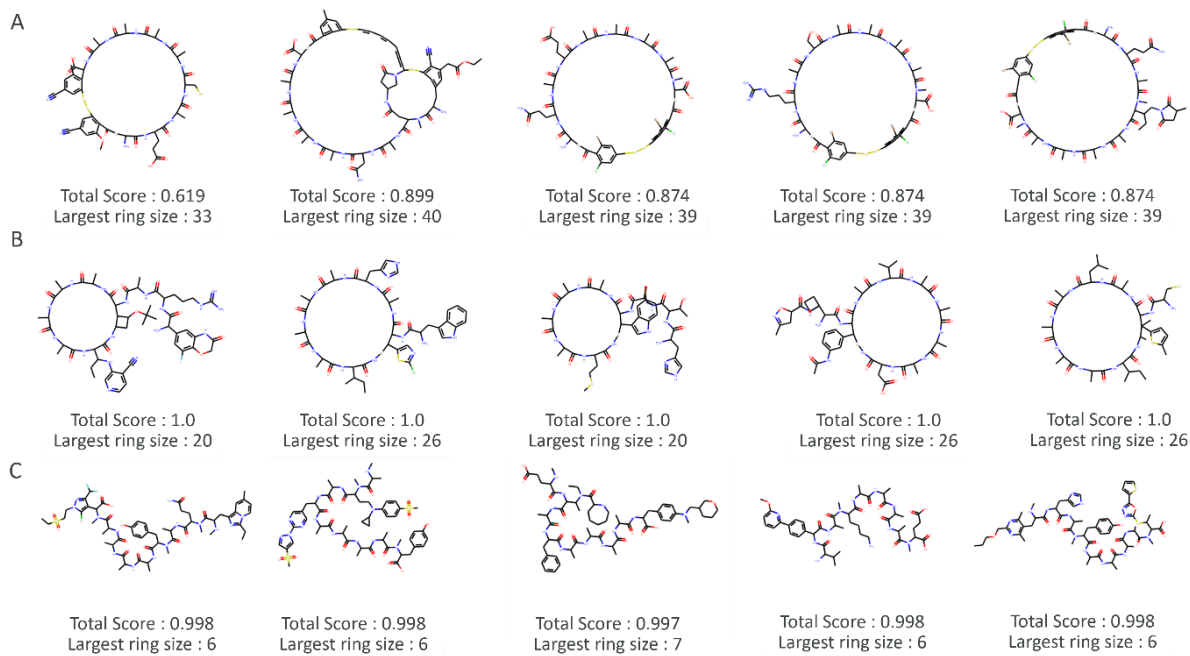

**Supplementary Figure 4.** Example peptides extracted from the last 10 steps of the RL runs that led to A) maximizing the ring size, B) preferring macrocycles with an upper limit of ring size, and C) minimizing the ring size, labelled with the total score and the largest ring size.

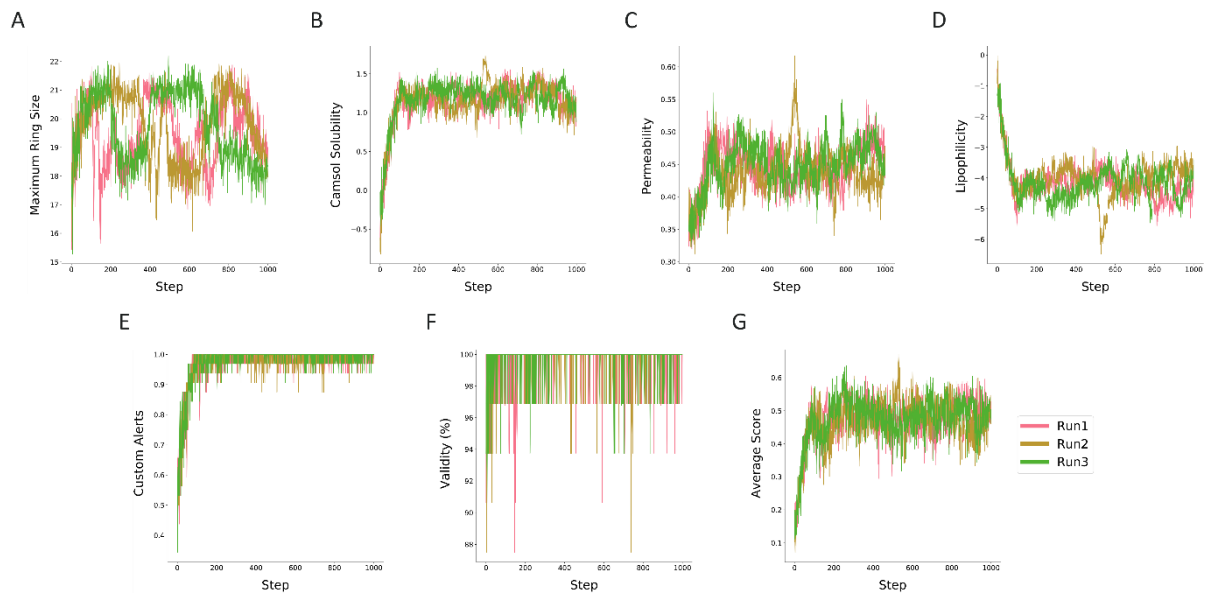

**Supplementary Figure 5.** The scoring components and the learning progress of the RL runs for Multi-Parameter Objective (MPO) task described in Scenario 2: Generating soluble and permeable cyclic peptides. The MPO task with the aim of designing soluble and permeable cyclic peptides were conducted in triplicate runs with multinomial sampling. The learning progresses of each triplicate were illustrated as A) the largest ring size, B) CAMSOL-PTM solubility score, C) the probability of belonging to the permeable class by the permeability predictor, D) the calculated lipophilicity score, E) the custom alerts score, F) the validity of the generated peptides and, G) the average score of the batch composed of all the scoring components of the MPO task.